\global\def\draftcontrol{0}

   \def\versionno{ n2quasinormal }

\catcode`\@=11

\expandafter\ifx\csname draftcontrol\endcsname\relax\global\def\draftcontrol{0}
\fi

{\count255=\time\divide\count255 by 60
\xdef\hourmin{\number\count255}
\multiply\count255 by-60\advance\count255 by\time
\xdef\hourmin{\hourmin:\ifnum\count255<10 0\fi\the\count255}}
\def\draftdate{\number\month/\number\day/\number\year\ \ \ \hourmin }

\newcommand\makepapertitle{\par
  \begingroup
    \renewcommand\thefootnote{\@fnsymbol\c@footnote}%
    \def\@makefnmark{\rlap{\@textsuperscript{\normalfont\@thefnmark}}}%
    \long\def\@makefntext##1{\parindent 1em\noindent
            \hb@xt@1.8em{%
                \hss\@textsuperscript{\normalfont\@thefnmark}}##1}%
     \newpage
     \global\@topnum\z@   
     \@makepapertitle
     \thispagestyle{empty}\@thanks
  \endgroup
  \setcounter{footnote}{0}%
  \global\let\thanks\relax
  \global\let\makepapertitle\relax
  \global\let\@makepapertitle\relax
  \global\let\@thanks\@empty
  \global\let\@author\@empty
  \global\let\@date\@empty
  \global\let\@title\@empty
  \global\let\title\relax
  \global\let\author\relax
  \global\let\date\relax
  \global\let\and\relax
  \def\version{\let\version\@version\@gobble}
}
\def\@makepapertitle{%
  \newpage
   \ifnum\draftcontrol=1 {}
   \version\versionno
   \vskip 3em%
   \else
   \hfill\hbox to 3cm {\parbox{4cm}{\@pubnum}\hss}%
   \vskip 3em%
   \fi
   \begin{center}%
   \let \footnote \thanks
     {\LARGE {\@title}}%
     \vskip 1.5em%
     {\normalsize
       \lineskip .5em%
       \begin{tabular}[t]{c}%
         \@author
       \end{tabular}\par}%
     \vskip 1.5em%
     {\@bstract}%
     \end{center}%
     \vskip 1.5em
     \@date%
   \par
}

\gdef\@pubnum{}
\def\pubnum#1{%
  \gdef\@pubnum{#1}}

\gdef\@bstract{}
\def\Abstract#1{%
  \gdef\@bstract{%
   \parbox{\textwidth-0pc}{%
   \centerline{\bf Abstract}\penalty1000%
\kern.2cm%
\noindent
\renewcommand\baselinestretch{1.0}%
{#1}}}
}

\def\ps@paper{\let\@mkboth\@gobbletwo%
     \ifnum\draftcontrol=1
    \def\@oddfoot{\hbox to \textwidth{\tiny \versionno \hfil\tiny\draftdate}%
    \hskip -\textwidth \hbox to \textwidth{\hfil\rm\thepage\hfil}}%
     \else\def\@oddfoot{\hbox to \textwidth{\hfil\rm\thepage\hfil}}
     \fi
     \let\@evenfoot\@oddfoot
}

\def\body{\clearpage
          \pagestyle{paper}
    }

\def\@version#1{\ifnum\draftcontrol=1
\typeout{}\typeout{#1}\typeout{}
\vskip3mm\centerline{\hbox{\fbox{\normalsize{\tt DRAFT -- #1 -- }
                   {\draftdate}}}}\vskip3mm
\fi}
\let\version\@version
\long\def\eqlabel#1{\ifnum\draftcontrol=1
                    \tag@false  
                    \tag*{(\theequation) \hbox to -0.2cm{\hspace{0cm}\small{#1}\hss}}
                    \refstepcounter{equation}
                    \edef\@currentlabel{\theequation}
                    \ltx@label{#1}          
                    \else
                    \label{#1}
                    \fi
                    }
\let\st@bibitem\@bibitem
\let\st@lbibitem\@lbibitem
\ifnum\draftcontrol=1
  \def\@bibitem#1{%
    \st@bibitem{#1}\a@@label{#1}\ignorespaces}
  \def\@lbibitem[#1]#2{%
    \st@lbibitem[#1]{#2}\a@@label{#2}\ignorespaces}
  \def\a@@label#1{%
    \gdef\a@lab{\smash{\normalfont\small#1}}
    \ifvmode
      \if@inlabel
        \global\setbox\@labels\hbox{%
          \llap{\a@lab\let\a@lab\relax
                \kern\@totalleftmargin\kern\marginparsep}%
          \box\@labels}%
      \fi
    \fi}
\fi

\documentclass[12pt,letterpaper]{article}

\usepackage{amsmath,amssymb,array,calc,epsfig,rotating,bm}
\usepackage[sort]{cite}
\usepackage{graphicx}
\usepackage{psfrag,verbatim}

\ifnum\draftcontrol=1
\fi

\renewcommand\baselinestretch{1.25}
\renewcommand\section{\@startsection {section}{1}{\z@}%
                                   {-3.5ex \@plus -1ex \@minus -.2ex}%
                                   {2.3ex \@plus.2ex}%
                                   {\normalfont\large\bfseries}}
\renewcommand\subsection{\@startsection{subsection}{2}{\z@}%
                                   {-3.25ex\@plus -1ex \@minus -.2ex}%
                                   {1.5ex \@plus .2ex}%
                                   {\normalfont\normalsize\bfseries}}
\renewcommand\subsubsection{\@startsection{subsubsection}{3}{\z@}%
                                   {-3.25ex\@plus -1ex \@minus -.2ex}%
                                   {1.5ex \@plus .2ex}%
                                   {\normalfont\normalsize\it}}
\renewcommand\paragraph{\@startsection{paragraph}{4}{\z@}%
                                   {-3.25ex\@plus -1ex \@minus -.2ex}%
                                   {1.5ex \@plus .2ex}%
                                   {\normalfont\normalsize\bf}}

\numberwithin{equation}{section}

\def\revise#1       {\raisebox{-0em}{\rule{3pt}{1em}}%
                     \marginpar{\raisebox{.5em}{\vrule width3pt\
                     \vrule width0pt height 0pt depth0.5em
                     \hbox to 0cm{\hspace{0cm}{%
                     \parbox[t]{4em}{\raggedright\footnotesize{#1}}}\hss}}}}

\newcommand\nxt[1]  {\\\fnxt#1}
\newcommand{\ie}{{\it i.e.,}\ }

\def\call         {{\cal L}}
\def\calm         {{\cal M}}
\def\caln         {{\cal N}}
\def\calo         {{\cal O}}
\def\calp         {{\cal P}}

\def\calv         {{\cal V}}

\def\complex      {{\mathbb C}}

\def\del          {\partial}

\def\Im           {{\rm Im\hskip0.1em}}

\def\sqr#1#2{{\vcenter{\vbox{\hrule height.#2pt
 \hbox{\vrule width.#2pt height#1pt \kern#1pt
 \vrule width.#2pt}\hrule height.#2pt}}}}

\newcommand{\ft}[2]{{\textstyle{\frac{#1}{#2}}}}

\newcommand{\kk}{\mathfrak{q}}
\newcommand{\ww}{\mathfrak{w}}

\def\a{\alpha}

\def\w{\omega}
\def\r{\rho}
\def\dd{\delta}

\def\aa1{\phi}
\def\cc1{\psi}

\def\l{\lambda}

\catcode`\@=12

\begin{document}


\title{\bf   On consistent truncations in $\caln=2^*$ holography}
\pubnum{UWO-TH-13/16}

\date{November 19, 2013}

\author{
Venkat Balasubramanian$^{1}$ and Alex Buchel$^{1,2}$\\[0.4cm]
\it $^1$\,Department of Applied Mathematics\\
\it University of Western Ontario\\
\it London, Ontario N6A 5B7, Canada\\
\it $^2$\,Perimeter Institute for Theoretical Physics\\
\it Waterloo, Ontario N2J 2W9, Canada
}

\Abstract{Although Pilch-Warner (PW) gravitational renormalization group
flow \cite{pw} passes a number of important consistency checks to be
identified as a holographic dual to a large-$N$ $SU(N)$ $\caln=2^*$
supersymmetric gauge theory, it fails to reproduce the free energy of
the theory on $S^4$, computed with the localization techniques.  This
disagreement points to the existence of a larger dual gravitational
consistent truncation, which in the gauge theory flat-space limit
reduces to a PW flow.  Such truncation was recently identified by
Bobev-Elvang-Freedman-Pufu (BEFP) \cite{befp}.  Additional bulk
scalars of the BEFP gravitation truncation might lead to
destabilization of the finite-temperature deformed PW flows, and thus
modify the low-temperature thermodynamics and hydrodynamics of
$\caln=2^*$ plasma. We compute the quasinormal spectrum of these bulk
scalar fields in the thermal PW flows and demonstrate that these modes
do not condense, as long as the masses of the $\caln=2^*$
hypermultiplet components are real.    
}

\makepapertitle

\body

\version\versionno
\tableofcontents

\section{Introduction and summary}\label{intro}
In \cite{pw} Pilch and Warner (PW) proposed a holographic renormalization group 
flow dual to $\caln=2$ supersymmetric $SU(N)$ gauge theory, obtained by 
turning on a mass term for the $\caln=2$ hypermultiplet of the parental 
$\caln=4$ $SU(N)$ supersymmetric Yang-Mills theory. This massive deformation of $\caln=4$ SYM 
is commonly referred to as $\caln=2^*$ theory.
As usual in most examples 
of gauge/gravity correspondence, the PW gravitational description is valid when
the gauge theory is in the planar limit, and has large (strictly speaking infinitely large) 
't Hooft coupling in the ultraviolet fixed point. While $\caln=2$ supersymmetric 
gauge theories have moduli space of vacua, the PW flow describes 
$\caln=2^*$ gauge theory at a particular point on a Coulomb branch \cite{bpp}.
Specifically, if $\Phi$ is an adjoint chiral multiplet (part of the $\caln=2$ vector multiplet),
its  Cartan subalgebra expectation values parameterize a generic Coulomb branch vacuum
\begin{equation}
\Phi={\rm diag}(a_1,a_2,\cdots,a_N)\,,\qquad \sum_{i} a_i=0\,,\qquad a_i\in \complex\,.
\eqlabel{cvacuum}
\end{equation}
In the planar limit, $N\to \infty$ and $g_{YM}^2\to 0$ with $g_{YM}^2 N$ kept fixed, it is natural to characterize the eigenvalue set \eqref{cvacuum}
with a continuous density distribution, $\r(a)$,
\begin{equation}
\int\int_{\complex} d^2a\ 
\r(a)=N\,.
\eqlabel{density}
\end{equation}
In \cite{bpp} the PW vacuum was shown to be uniquely identified with the following eigenvalue distribution:
\begin{equation}
\begin{split}
&\Im(a_i)=0\,,\qquad a_i\in [-a_0,a_0]\,,\qquad a_0^2=\frac{m^2 g_{YM}^2 N}{4\pi^2}\,,\\
&\r(a)=\frac{8\pi}{m^2 g_{YM}^2}\ \sqrt{a_0^2-a^2}\,,\qquad \int_{-a_0}^{a_0}da\ 
\r(a)=N\,,
\end{split}
\eqlabel{pwdistr}
\end{equation}
where $m$ is the hypermultiplet mass, and  $\l\equiv g_{YM}^2 N\gg 1$ is a 't Hooft coupling 
of the parental $\caln=4$ SYM.

The question {\it ``What makes the PW vacuum \eqref{pwdistr} special?''} was answered in 
\cite{Buchel:2013id}: turns out, supersymmetric formulation of $\caln=2^*$ gauge theory 
on $S^4$, in the planar limit and at large 't Hooft coupling, 
lifts all the Coulomb branch moduli, except for a single point; this single point, in the 
$S^4$ decompactification limit is precisely the vacuum 
\eqref{pwdistr}.
Furthermore, it was shown in \cite{Buchel:2013id} that the 
expectation value of the supersymmetric Wilson loop in PW geometry correctly reproduces the
decompactification limit of the corresponding large-$N$ matrix model \cite{pestun} computation.   

However, the agreement between the matrix model and the holographic computations
does not extend to the free energy of the $\caln=2^*$ theory on $S^4$ \cite{bloc}. 
Although a holographic renormalization produces a scheme-dependent result for the 
free energy\footnote{The scheme dependence arises via the finite counterterms 
in the holographic renormalization.}, it is possible to completely parameterize 
these free energy ambiguities \cite{n2hydro,Buchel:2012gw}. One can prove then that there does not exist a 
choice of a scheme in which the free energy of the $S^4$-compactified PW flow would 
agree with the free energy of the $\caln=2^*$ theory computed in the large-$N$, 
and large $\l$,
saddle point of the corresponding matrix integral \cite{bloc}.      
It was suggested in \cite{bloc} (also in \cite{btalk}) that this free energy puzzle
points to the existence of an ``enlarged'' holographic truncation of the $\caln=2^*$
gauge theory, which in addition to PW bulk scalars $\a$ and $\chi$ contains additional 
scalars representing the coupling of the gauge theory to $S^4$ background metric, necessary to 
insure the curved-space supersymmetry \cite{pestun}. This alternative gravitational 
truncation was found in \cite{befp} (BEFP). At this stage the full ten-dimensional uplift of the 
BEFP effective action is unknown; nonetheless, BEFP five-dimensional action is enough to demonstrate that 
holographic free energy (in a certain renormalization scheme) agrees precisely with the 
matrix model localization result.         

PW effective action, as a holographic dual to $\caln=2^*$ supersymmetric gauge theory, has been 
extensively used as a benchmark for gravitational computations of  the thermodynamics \cite{bl,bdkl,yaffe},
the hydrodynamics \cite{bbs,bbulk1,bbulk2,relax}, and the entanglement entropy 
\cite{ent} of strongly coupled nonconformal gauge theory plasmas. For example,
in \cite{n2crit} a critical phenomena in $\caln=2^*$ plasma was identified, which appears 
to be  outside the dynamical universality classes established by Hohenberg and Halperin \cite{hh}. 
Existence of the alternative gravitational dual identified by BEFP
raises the question as to what properties of the strongly coupled $\caln=2^*$ plasma can be reliably 
computed with PW effective action. From the gravitation perspective, the question is whether 
black hole solutions found in the framework of PW effective action are stable with respect to 
(normalizable) fluctuations of the additional BEFP bulk scalars. 

In this paper we compute the spectrum of quasinormal modes of BEFP bulk scalars in PW black brane geometries.
We show that as long as the masses of the hypermultiplet components are real, \ie both 
the bosonic component $m_b^2$, and the fermionic components $m_f^2$, of the hypermultiplet mass-squared are positive,
all quasinormal modes attenuate --- the PW horizon bulk geometries are stable. Thus, the 
low-energy properties of $\caln=2^*$ plasma can be reliably computed from the holographic PW action.  
On the other hand, we find that there exist a regime, with tachyonic hypermultiplet
component masses, where $SU(2)_V\subset SU(4)_R$ symmetry (see \cite{befp}) is spontaneously broken
at sufficiently low temperatures. 
 
The rest of the paper is organized as follows. In section \ref{action} we compare BEFP and PW effective 
gravitational actions representing holographic dual to $\caln=2^*$ gauge theory at strong coupling. 
In section \ref{quasinormal} we derive equations of motion and specify appropriate boundary conditions 
for BEFP normalizable fluctuations about thermal PW backgrounds. We compute the corresponding quasinormal spectra 
in two regimes:
\nxt ``bosonic'' $\caln=4$ deformations: $m_b^2\ne 0\,,\ m_f^2=0$;
\nxt "supersymmetric'' $\caln=4$ deformations: $m_b^2=m_f^2\equiv m^2$. \\
We find that while the thermal PW backgrounds are stable with respect to BEFP fluctuations 
for $m_b^2>0$ and $m^2>0$, the  $SU(2)_V$ symmetry breaking fluctuations destabilize 
finite temperature PW flows with negative $m_b^2$ and $m^2$ at low temperatures. 
We conclude in section \ref{conclude}.

\section{PW versus BEFP effective actions}\label{action}
We begin with description of the PW effective action \cite{pw}, and its gravitational RG flows 
dual to $\caln=2^*$ plasma at strong coupling \cite{bl,bdkl}. 
The action of the effective five-dimensional gauged supergravity including the
scalars $\alpha$ and $\chi$ (dual to mass terms for the bosonic and
fermionic components of the hypermultiplet respectively) is given by
\begin{equation}
\begin{split}
S=&\,
\int_{\calm_5} d\xi^5 \sqrt{-g}\ \call_{PW}\\
=&\frac{1}{4\pi G_5}\,
\int_{\calm_5} d\xi^5 \sqrt{-g}\left[\ft14 R-3 (\del\a)^2-(\del\chi)^2-
\calp\right]\,,
\end{split}
\eqlabel{action5}
\end{equation}
where the potential%
\footnote{We set the five-dimensional gauged
supergravity coupling to one. This corresponds to setting the
radius $L$ of the five-dimensional sphere in the undeformed metric
to $2$.}
\begin{equation}
\calp=\frac{1}{16}\left[\frac 13 \left(\frac{\del W}{\del
\a}\right)^2+ \left(\frac{\del W}{\del \chi}\right)^2\right]-\frac
13 W^2\,,
 \eqlabel{pp}
\end{equation}
is a function of $\alpha$ and $\chi$, and is determined by the
superpotential
\begin{equation}
W=- e^{-2\alpha} - \frac{1}{2} e^{4\alpha} \cosh(2\chi)\,.
\eqlabel{supp}
\end{equation}
In our conventions, the five-dimensional Newton's constant is
\begin{equation}
G_5\equiv \frac{G_{10}}{2^5\ {\rm vol}_{S^5}}=\frac{4\pi}{N^2}\,.
\eqlabel{g5}
\end{equation}
Regular horizon black brane solutions of \eqref{action5} are found 
within the background ansatz:
\begin{equation}
\begin{split}
&ds_5^2=(2x-x^2)^{-1/2} e^{2a}\biggl(-(1-x)^2 (dt)^2+(d\vec{x})^2\biggr)+g_{xx} (dx)^2\,,\\ 
&\r_6\equiv e^{6\a}\,,\qquad c\equiv \cosh 2\chi\,,
\end{split}
\eqlabel{backpw}
\end{equation}
where $\{a,\r_6,c\}$ are functions of a radial coordinate 
\begin{equation}
x\in [0,1)\,.
\eqlabel{defx}
\end{equation}
The physical parameters of the plasma are encoded in the asymptotic coefficients 
of the metric and the bulk scalar functions near the boundary ($x\to 0_+$) and 
near the horizon ($x\to 1_-$). Specifically, as $x\to 0_+$ we have 
\begin{equation}
\begin{split}
&\r_6=1+6 x^{1/2} \left(\r_{1,0}+\r_{1,1}\ln x\right)+\calo(x\ln^2 x)\,,\\
&c=1+x^{1/2} c_{1,0}+x \left(c_{2,0}+\frac 13 c_{1,0}^2\ln x\right)+\calo(x^{3/2}\ln^2 x)\,,\\
&a=-\frac{1}{18} x^{1/2} c_{1,0}+\calo(x\ln^2 x)\,,
\end{split}
\eqlabel{uv}
\end{equation} 
and as $y\equiv (1-x)\to 0_+$ we have 
\begin{equation}
\r_6=\r_0+\calo(y^2)\,,\qquad c=c_0+\calo(y^2)\,,\qquad a=a_0+a_1 y^2+\calo(y^4)\,.
\eqlabel{ir}
\end{equation}
The  temperature and the hypermultiplet component masses are given by \cite{bdkl} 
\begin{equation}
\begin{split}
&(\pi T)^2=e^{2a_0}\ \frac{4+\r_0^2+8\r_0c_0-\r_0^2c_0^2}{48(1+4a_1)\r_0^{2/3}}\,,\\
&\left(\frac{m_b}{\pi T}\right)^2=12\sqrt{2}\ e^{6a_0} \r_{1,1}\,,\qquad 
\left(\frac{m_f}{\pi T}\right)^2=\sqrt{2}\ e^{6a_0} c_{1,0}\,.
\end{split}
\eqlabel{physical}
\end{equation}
Note that in \eqref{physical} 
we set $2\pi T=1$ in the conformal case, \ie when  $\r_6(x)=c(x)\equiv 1, a(x)\equiv0$. 
This does not affect final results 
as long as we express them in dimensionless ratios (the temperature is one of the three microscopic 
mass scales in $\caln=2^*$ plasma, and for dimensionless results we can set one of these 
scales to an arbitrary value). The full $T$ dependence can be restored via an arbitrary shift of 
the metric factor $a$.  Furthermore, from \eqref{physical}, 
the supersymmetry condition, \ie $m_b^2=m_f^2$, constraints 
\begin{equation}
c_{1,0}=12\r_{1,1}\,.
\eqlabel{susyconst}
\end{equation}

It was argued in \cite{bloc,btalk} that PW effective action \eqref{action5} can not be the consistent 
gravitational truncation\footnote{It is a consistent truncation of 
type IIB supergravity.} of the holographic dual to $\caln=2^*$ plasma --- it fails to reproduce the 
exact field-theoretic computations of the free energy of the theory on $S^4$. 
The correct gravitational dual was found in \cite{befp}. The BEFP effective action is 
given by 
\begin{equation}
\begin{split}
S_{BEFP}=&\,
\int_{\calm_5} d\xi^5 \sqrt{-g}\ \call_{BEFP}\\
=&\frac{1}{4\pi G_5}\,
\int_{\calm_5} d\xi^5 \sqrt{-g}\left[ R-12 \frac{(\del\eta)^2}{\eta^2}
-4 \frac{(\del{\vec X})^2}{(1-\vec{X}^2)^2}
-\calv\right]\,,
\end{split}
\eqlabel{befp}
\end{equation}
with the potential 
\begin{equation}
\calv=-\left[\frac{1}{\eta^4}+2\eta^2\ \frac{1+\vec{X}^2}{1-\vec{X}^2}
-\eta^8\ \frac{(X_1)^2+(X_2)^2}{(1-\vec{X}^2)^2}
\right]\,,
 \eqlabel{pbefp}
\end{equation}
where $\vec{X}=\left(X_1,X_2,X_3,X_4,X_5\right)$ are five of the scalars and $\eta$ is the sixth. 
The symmetry of the action reflects the symmetries of the dual gauge theory \cite{befp}: the 
two scalars $(X_1,X_2)$ form a doublet under the $U(1)_R$ part of the gauge group, 
while $(X_3,X_4,X_5)$ form a triplet under $SU(2)_V$ and $\eta$ is neutral.  
The PW effective action is recovered as a consistent truncation of \eqref{befp} with 
\begin{equation}
X_2=X_3=X_4=X_5=0\,,
\eqlabel{truncate}
\end{equation}
provided we identify the remaining BEFP scalars $(\eta,X_1)$ with the PW scalars $(\a,\chi)$ as follows
\begin{equation}
e^\a\equiv \eta\,,\qquad \cosh 2\chi =\frac{1+(X_1)^2}{1-(X_1)^2} \,.
\eqlabel{id}
\end{equation}
Note that once $m_f\ne 0$ (correspondingly $X_1\ne 0$), the $U(1)_R$ symmetry is
explicitly broken; on the contrary, $SU(2)_V$ remains unbroken in truncation to PW.   

The goal of this paper is to study stability of gravitational solutions obtained in 
PW truncation of BEFP holographic dual to $\caln=2^*$ gauge theory. 
The effective action describing fluctuations of PW backgrounds within BEFP is obtained 
linearizing \eqref{befp} in $X_2,X_3,X_4,X_5$ scalar fields:
\begin{equation}
\begin{split}
&\dd \call\equiv  \call_{BEFP}-\call_{PW}+\calo(X_i^4)\equiv  \dd\call_2+\dd\call_V\,,\\
&\dd\call_2=-(1+c)^2 (\del X_2)^2-\frac {1+c}{4}\left((c^2+c) \r_6^{4/3}-4 (1+c) \r_6^{1/3}
+\frac{4(\del c)^2}{c^2-1}\right) (X_2)^2\,,\\ 
&\dd\call_V=-(1+c)^2 (\del \vec{X}_V)^2-\frac {1+c}{4}\left((c^2-1) \r_6^{4/3}-4 (1+c) \r_6^{1/3}
+\frac{4(\del c)^2}{c^2-1}\right) (\vec{X}_V)^2\,, 
\end{split}
\eqlabel{linear} 
\end{equation}
where $\vec{X}_V=(X_3,X_4,X_5)$. Note that $\dd \call $ is $SU(2)_V$ invariant; as a result it is enough to 
consider a spectrum of only one of $\vec{X}_V$ components. In what follows we choose the latter to be 
$X_3$.

\section{Spontaneous breaking of $SU(2)_V$ symmetry in thermal PW flows}\label{quasinormal}
In this section we study the spectrum of normalizable fluctuations of \eqref{linear} in thermal PW 
backgrounds \eqref{backpw}-\eqref{physical}. The framework for such analysis is equivalent to the 
one of the chiral symmetry breaking quasinormal modes in cascading gauge theory plasma developed 
in \cite{bchiral}. As mentioned in the introduction, we focus on bosonic ($m_f^2=0$) and supersymmetric 
$m_b^2=m_f^2$ thermal flows --- extensions to more general flows are straightforward, and we do not expect
physically different results compared to the ones reported here.  

We omit subscript for $X$ whenever the discussion is identical for $X_2$ and $X_3$ BEFP scalars. 
Without the loss of generality we assume
\begin{equation}
X(t,\vec{x},x)=e^{-i \w t+ i k x_3} F(x)\,.
\eqlabel{fluc}
\end{equation}
The radial wave-function $F(x)$ satisfies a homogeneous second order differential 
equation\footnote{These equations are too long to be presented here. They are available 
as supplement to the {\it arXiv.org} submission of this paper.}.
The physical fluctuations must satisfy an incoming wave boundary 
condition at the PW black brane horizon, and be normalizable 
at the asymptotic $x\to 0_+$ boundary. Introducing 
\begin{equation}
\ww=\frac{\w}{2\pi T}\,,\qquad \kk=\frac{k}{2\pi T}\,,
\eqlabel{defwq}
\end{equation}
the former condition implies 
\begin{equation}
F(x)= (1-x)^{-i\ww} f(x)\,,
\eqlabel{inc}
\end{equation}
with $f(x)$ being regular at the horizon, \ie as $x\to 1_-$. The equation 
of motion for $f(x)$ is complex --- it becomes real once once we 
introduce 
\begin{equation}
\ww=-i \Omega\,,\qquad \Im(\Omega)=0\,.
\eqlabel{realw}
\end{equation}
Using the background asymptotic expansion \eqref{uv}, the normalizability 
condition for $f$ at the $x\to 0_+$ boundary translates into the following 
asymptotic solutions,
\nxt for the bosonic thermal flows:
\begin{equation}
\begin{split}
&f_2=f_{2,1,0}\ x^{3/4}\ \left(1+\frac{x^{1/2}}{\sqrt{2}}
\left((2\pi T\kk)^2+ (2\pi T\Omega)^2 \right)
+\calo(x \ln^2 x)
\right)\,,\\
&f_3=f_{3,1,0}\ x^{1/2}\ \biggl(1+x^{1/2}\left(\sqrt{2}(2\pi T\kk)^2+\sqrt{2}(2\pi T\Omega)^2
+8\r_{1,1}-2\r_{1,0}-2\r_{1,1}\ln x\right)\\
&+\calo(x \ln^2 x)\biggr)\,;
\end{split}
\eqlabel{fluvb}
\end{equation}
\nxt  for the supersymmetric thermal flows:
\begin{equation}
\begin{split}
&f_2=f_{2,1,0}\ x^{3/4}\ \left(1+\frac{x^{1/2}}{\sqrt{2}}
\left((2\pi T\kk)^2+ (2\pi T\Omega)^2 -4\sqrt{2}\r_{1,1} \right)
+\calo(x \ln^2 x)
\right)\,,\\
&f_3=f_{3,1,0}\ x^{1/2}\ \biggl(1+x^{1/2}\left(\sqrt{2}(2\pi T\kk)^2+\sqrt{2}(2\pi T\Omega)^2
+\frac{22}{3}\r_{1,1}-2\r_{1,0}-2\r_{1,1}\ln x\right)\\
&+\calo(x \ln^2 x)\biggr)\,.
\end{split}
\eqlabel{fluvs}
\end{equation}
Since the equation of motion for $f$ is homogeneous,
without the loss of generality we can set $f(1)=1$. The IR, \ie as $y\equiv (1-x)
\to 0_+$, 
asymptotic expansion then takes form
\begin{equation}
f=1+\calo(y^2)\,.
\eqlabel{irasymptotic}
\end{equation}
Notice that for each mode ($f_2$ or $f_3$) we have a single adjustable parameter: $f_{2,1,0}$ 
or $f_{3,1,0}$, in order to solve a boundary value problem for a 
second-order differential equations for $f$.
As a result, a solution produces a dispersion relation for the BEFP fluctuations:
\begin{equation}
\Omega=\Omega(\kk^2)\,.
\eqlabel{disp}
\end{equation}
The quasinormal modes signal an instability in plasma provided 
\begin{equation}
\Im(\ww)>0\ \Leftrightarrow\ \Omega<0\,,\qquad {\rm provided}\qquad \Im(\kk)=0\,.
\eqlabel{instablity}
\end{equation}

\begin{figure}[t]
\begin{center}
\psfrag{mt}{{$\frac{m_b^2}{T^2}$}}
\psfrag{q}{{$\kk^2$}}
\includegraphics[width=3in]{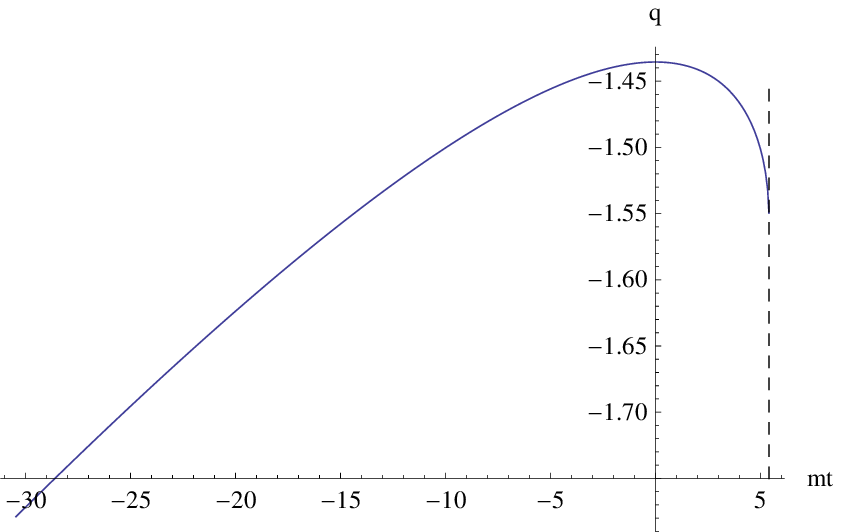}
\includegraphics[width=3in]{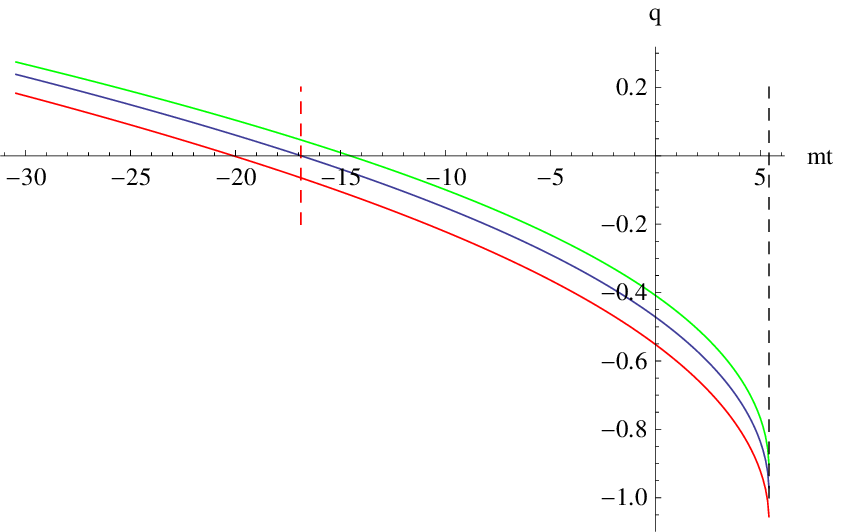}
\end{center}
  \caption{(Colour online) {\bf Left panel:} dispersion relation of the BEFP fluctuations $f_2$ 
in the bosonic thermal PW flows as a function of $\frac{m_b^2}{T^2}$ 
at the threshold of instability: $(\ww=0,\kk^2)$.
{\bf Right panel:} dispersion relation of the BEFP fluctuations $f_3$ 
in the bosonic thermal PW flows.
The solid blue line represents the dispersion relation  
at the threshold of instability: $(\ww=0,\kk^2)$.
The solid green line indicates 
quasinormal modes with $(\Omega=0.1, \kk^2)$ as a function of $\frac{m_b^2}{T^2}$;
the solid red  line indicates 
quasinormal modes with $(\Omega=-0.1, \kk^2)$ as a function of $\frac{m_b^2}{T^2}$.
The red dashed vertical line indicates the onset of instability: for smaller values of  
$\frac{m_b^2}{T^2}$ the $f_3$ mode condenses with spontaneous breaking of 
$SU(2)_V$ symmetry of thermal bosonic PW flows. 
In {\bf both panels}, the black dashed vertical lines represent the 
critical point in $\caln=2^*$ phase diagram, 
see \cite{n2crit}. 
 } \label{figure1}
\end{figure}

\begin{figure}[t]
\begin{center}
\psfrag{mt}{{$\frac{m^2}{T^2}$}}
\psfrag{q}{{$\kk^2$}}
\includegraphics[width=3in]{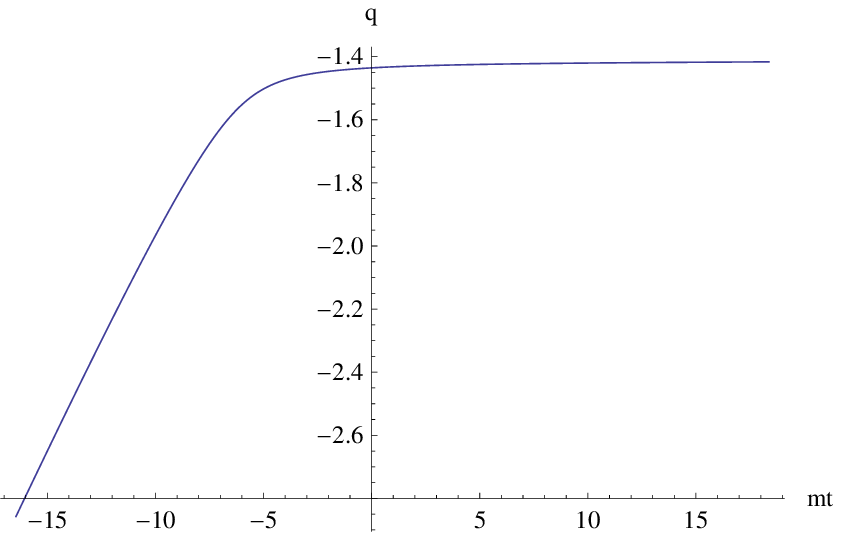}
\includegraphics[width=3in]{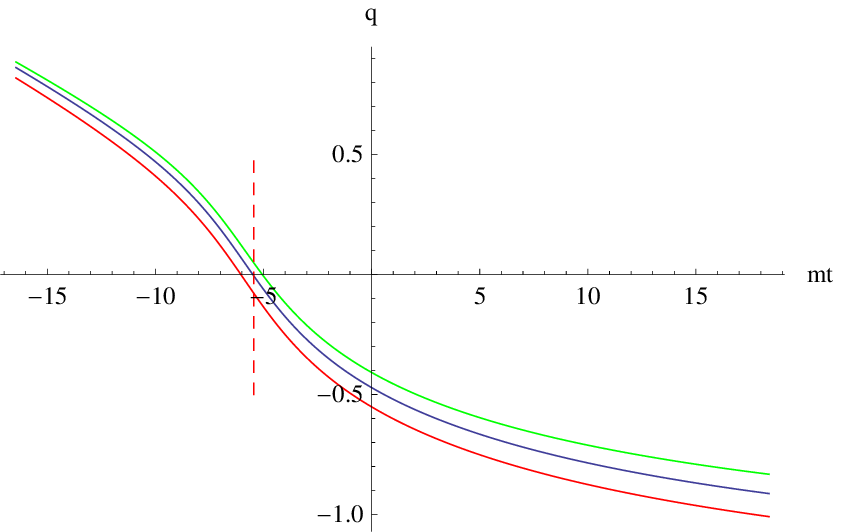}
\end{center}
  \caption{(Colour online) {\bf Left panel:} dispersion relation of the BEFP fluctuations $f_2$ 
in the supersymmetric thermal PW flows as a function of $\frac{m^2}{T^2}$ 
at the threshold of instability: $(\ww=0,\kk^2)$.
{\bf Right panel:} dispersion relation of the BEFP fluctuations $f_3$ 
in the supersymmetric thermal PW flows.
The solid blue line represents the dispersion relation  
at the threshold of instability: $(\ww=0,\kk^2)$.
The solid green line indicates 
quasinormal modes with $(\Omega=0.1, \kk^2)$ as a function of $\frac{m^2}{T^2}$;
the solid red  line indicates 
quasinormal modes with $(\Omega=-0.1, \kk^2)$ as a function of $\frac{m^2}{T^2}$.
The red dashed vertical line indicates the onset of instability: for smaller values of  
$\frac{m^2}{T^2}$ the $f_3$ mode condenses with spontaneous breaking of 
$SU(2)_V$ symmetry of thermal supersymmetric PW flows. 
 } \label{figure2}
\end{figure}

The results of the analysis of the dispersion relation of BEFP fluctuations 
are presented in Figures~\ref{figure1} (for bosonic thermal flows) and \ref{figure2} 
(for supersymmetric thermal flows).
In principle, we expect discrete branches of the quasinormal modes
distinguished by the number of nodes in radial profiles 
$f$. In what follows we consider 
only the lowest quasinormal mode, which has monotonic radial profile.

The black dashed vertical lines in Fig.~\ref{figure1} represent the critical 
temperature $T_c$ in $\caln=2^*$ plasma (see \cite{n2crit} for detailed discussion), 
\begin{equation}
\frac{m_b}{T}\approx 2.32591\,.
\eqlabel{tcrit}
\end{equation}
The left panel presents the spectrum of $f_2$ fluctuations, and the right panel 
presents the spectrum of $f_3$ fluctuations. 
Solid blue lines  indicate dispersion relations for BEFP fluctuations 
at the threshold of instability, \ie with $\ww=0$. Notice that on-shell 
fluctuations of $f_2$ mode have dispersion with $\kk^2<0$, implying that they are massive.  
Thus,  BEFP mode $f_2$ never condenses in thermal bosonic PW flows. On the contrary,
BEFP mode $f_3$ condenses with spontaneous breaking of $SU(2)_V$ symmetry of the 
PW truncation for 
\begin{equation}
\frac{m_b^2}{T^2}< \frac{m_b^2}{T_{u,b}^2}\approx -16.8(9)\,,
\eqlabel{unstableb}
\end{equation}
indicated by the red dashed vertical line in the right panel. 
At a given temperature, quasinormal modes with $\kk^2$ below the 
momenta of the modes at the threshold of instability (blue line) 
are expected to have $\Omega<0$ (indicating a genuine tachyonic instability),
while modes with $\kk^2$ above the momenta of the modes at the threshold of instability 
are expected to have $\Omega>0$ (indicating  stable excitations). This is precisely what we 
find:  the red line in the right panel have $\Omega=-0.1$ and the green line 
indicate quasinormal modes with $\Omega=0.1$.

The supersymmetric thermal PW flows exhibit identical pattern, see Fig.~\ref{figure2}:
BEFP mode $f_2$ is always stable, while BEFP mode $f_3$ 
condenses with spontaneous breaking of $SU(2)_V$ symmetry of the 
PW truncation for 
\begin{equation}
\frac{m^2}{T^2}< \frac{m^2}{T_{u,s}^2}\approx -5.4(5)\,.
\eqlabel{unstableb}
\end{equation}

\section{Conclusion}\label{conclude}
The BEFP construction \cite{befp} resolves puzzling feature of the $\caln=2^*$ holography: 
the disagreement between the holographic free energy computation of the $S^4$-compactified 
supersymmetric PW flows 
and the exact field theoretic computations of the free energy via localization.  
This is achieved by embedding the PW effective action as $SU(2)_V$-invariant 
sector of ``enlarged'' five-dimensional gauged supergravity. 
Compare to PW effective action, BEFP 
action contains a triplet of $SU(2)_V$ bulk scalar fields, and one additional neutral 
scalar. These additional bulk scalars are necessary to model the coupling of $\caln=2^*$ gauge 
theory to background $S^4$ metric, as required by the supersymmetry \cite{pestun}. 

In this paper we studied the stability of PW embedding within BEFP for thermal 
flows, previously used to study various thermodynamic and hydrodynamic 
properties of $\caln=2^*$ plasma. We demonstrated that the embedding is stable 
for all physical masses of the gauge theory hypermultiplet components. 
Interestingly, the $SU(2)_V$ symmetry can be spontaneously broken,
but this occurs for the tachyonic masses of the $\caln=2$ 
hypermultiplet components, and at sufficiently low temperatures.

~\\
\section*{Acknowledgments}
We would like to thank Nikolay Bobev for valuable discussions.
Research at Perimeter
Institute is supported by the Government of Canada through Industry
Canada and by the Province of Ontario through the Ministry of
Research \& Innovation. We gratefully acknowledge further support by an
NSERC Discovery grant.

\end{document}